\documentclass[reprent,aps,prb,twocolumn
]{revtex4-2}
\usepackage{amsmath,amssymb}
\usepackage[dvipdfmx]{graphicx}
\usepackage{bm}
\usepackage{braket}
\usepackage{color}

\newcounter{num}
\newcommand{\Rnum}[1]{\setcounter{num}{#1} \Roman{num}}

\begin{document}
\title{
Magnetic instability under ferroaxial moment
}

\author{Akane Inda and Satoru Hayami}
\affiliation{
  Faculty of Science, Hokkaido University, Sapporo 060-0810, Japan
}

\begin{abstract}
Magnetic anisotropy is one of the important factors in determining magnetic structures. 
A type of magnetic anisotropy is closely related to the symmetry of crystals. 
We theoretically investigate magnetic anisotropy and its related magnetic instability arising from an electric axial moment, which appears under the breaking of the mirror symmetry parallel to the moment but does not require the breakings of both spatial inversion and time-reversal symmetries.
By performing perturbation and mean-field calculations in a complementary way, we show the appearance of the in-plane magnetic anisotropy when the electric axial moment occurs, which tends to tilt in-plane spin moments from the crystal axes in collaboration with relativistic spin--orbit coupling. 
We demonstrate such a tendency for single-site and four-site cluster models, the latter of which leads to the instability toward a spin vortex phase accompanying magnetic monopole and magnetic toroidal dipole.
\end{abstract}

\maketitle
\section{Introduction}
The stability of magnetic structures has long been studied in lots of materials.
Depending on the types of magnetic interactions and anisotropy, noncollinear and noncoplanar magnetic structures as well as collinear ferromagnetic and antiferromagnetic structures can be realized. 
The appearance of magnetic anisotropy is often related to the symmetry of the crystal, where the interplay between the relativistic spin--orbit coupling (SOC) and crystalline electric field plays an important role. 
For example, the Dzyaloshinskii-Moriya (DM) interaction is present when the spatial inversion symmetry at the bond center is lost~\cite{Dzyaloshinsky1958jpcs_weak-ferromagnetism,moriya1960pr_superexchange}, which gives rise to a helical spiral state and skyrmion crystal~\cite{nagaosa2013topological}. 
Another example is the Kitaev-type exchange interaction that arises from the strong SOC for the discrete rotational symmetry~\cite{jackeli2009mott, winter2017models}, which induces noncoplanar spin textures~\cite{Michael_PhysRevB.91.155135, Lukas_PhysRevLett.117.277202, yao2016topological}. 
The relation between magnetic interactions and crystal symmetry has been so far classified in real space~\cite{kaplan1983single, Shekhtman_PhysRevB.47.174} and momentum space~\cite{Yambe_PhysRevB.106.174437}. 
The complicated magnetic textures induced by magnetic anisotropy lead to unconventional physical phenomena, such as the topological Hall effect under noncoplanar spin textures~\cite{Ye_PhysRevLett.83.3737, Ohgushi_PhysRevB.62.R6065, tatara2002chirality, Nagaosa_RevModPhys.82.1539} and nonlinear longitudinal/transverse transport under noncollinear/noncoplanar spin textures~\cite{tokura2018nonreciprocal, Xiao_PhysRevB.100.165422, Hayami_PhysRevB.106.014420, Kirikoshi_PhysRevB.107.155109}. 

In the present study, we investigate the origin and the role of magnetic anisotropy under an electric axial moment, whose uniform component is referred to as the ferroaxial (or ferrorotational) moment.
The ferroaxial moment corresponds to a time-reversal-even axial dipole moment, which appears when the mirror symmetry parallel to the moment is lost but remains spatial inversion ($\mathcal{P}$) and time-reversal ($\mathcal{T}$) symmetries~\cite{hlinka2016aps_Symmetry_Guide_to_Ferroaxial_Transitions}. 
The ordered state of such a ferroxial moment has been experimentally observed in materials like CaMn$_7$O$_{12}$~\cite{johnson2012prl_CaMn7O12}, RbFe(MoO$_4$)$_2$~\cite{jin2020natphys_RbFeMoO42,Hayashida2021PRM_ferroaxial_domain}, NiTiO$_3$~\cite{hayashida2020natcom_Visualization_of_ferroaxial_domains_in_an_order-disorder_type_ferroaxial_crystal,  yokota2022npj_three-dimensional_imaging}, Ca$_5$Ir$_3$O$_{12}$~\cite{Hasegawa_doi:10.7566/JPSJ.89.054602, hanate2021first, hayami2023cluster, hanate2023space}, and BaCoSiO$_4$~\cite{xu2022prb_BaCoSiO4}.
Although the ferroaxial moment does not directly couple to neither electric field nor magnetic field owing to the even parity in terms of the
$\mathcal{P}$ and $\mathcal{T}$ symmetries, recent studies clarified that it becomes the origin of rich transverse responses of the conjugate physical quantities~\cite{cheong2021permutable, Hayami2022jpsj_spincurrent} such as the spin current generation~\cite{Roy2022prm_spin-current, Hayami2022jpsj_spincurrent}, antisymmetric thermopolarization~\cite{nasu2022prb_thermopolarization}, nonlinear transverse magnetization~\cite{inda2023jpsj}, unconventional Hall effect~\cite{Hayami_PhysRevB.108.085124}, and nonlinear magnetostriction~\cite{kirikoshi2023rotational}. 
Meanwhile, magnetic instability under the ferroaxial ordering has not been fully clarified in spite of the Kramers degeneracy owing to the $\mathcal{T}$ symmetry. 
Thus, it is desired to examine what types of magnetic instabilities occur under the ferroaxial ordering. 
Especially, it is important to understand how magnetic anisotropy is generated by the onset of the ferroaxial ordering, which might be helpful for understanding and exploring magnetic phase transitions in ferroaxial materials. 

For that purpose, we analyze a typical $d$-orbital model with the $d^1$ configuration based on the multipole representation~\cite{hayami2018microscopic,Watanabe2018PRB_Symmetry-analysis, Watanabe2018prb_Group-theoretical-classification,Spaldin2008jpcm_toroidal-moment, Hlinka2014prl_eight-types,hayami2018prb_Classification_of_atomic-scale_multipoles, kusunose2020complete, yatsushiro2021prb_122, kusunose2022generalization}, where four types of multipoles with distinct $\mathcal{P}$ and $\mathcal{T}$ parities, electric, magnetic, magnetic toroidal, and electric toroidal, constitute a complete basis set in the low-energy Hilbert space~\cite{hayami2018microscopic,hayami2018prb_Classification_of_atomic-scale_multipoles, kusunose2020complete}.
Since the dipole component of the electric toroidal multipoles, i.e., the electric toroidal dipole (ETD), corresponds to the ferroaxial moment, we examine the magnetic instability in the presence of the ETD.  
First, we perform perturbation and mean-field calculations for the single-site $d$-orbital model. 
As a result, we show that the synergy between the molecular field arising from the ETD moment and the SOC leads to single-ion magnetic anisotropy, which tends to tilt the in-plane spin moments from the crystal axis. 

Then, we analyze the $d$-orbital model in a four-site cluster. 
We find that the stability of a vortex spin state accompanying both magnetic monopole and magnetic toroidal dipole is enhanced by the magnetic anisotropy characteristic of the ferroaxial moment. 
We show that the ratio of magnetic monopole and magnetic toroidal dipole becomes comparable to each other when the magnitude of the SOC is comparable to that of the ETD molecular field. 
Our results indicate that the magnetic anisotropy arising from the ferroaxial moment can be a source of intriguing magnetic phases, which might exhibit a variety of cross-correlation phenomena.

The remaining part of this paper is organized as follows: In Sec.~\ref{sec:anisotropy}, we briefly introduce the ferroaxial moments based on the multipole representation.
Then, we present a single-site $d$-orbital model, and we show the role of the ferroaxial moment on the magnetic anisotropy through the second-order perturbative analysis. 
We also numerically evaluate 
the magnetic anisotropy by performing the mean-field calculations. 
Then, we show the stable magnetic textures under ETD moments in a four-site tetragonal cluster within the mean-field approximation in Sec.~\ref{sec:instability}. 
We show that the SOC under the ETD moment leads to a spin vortex phase accompanying both the magnetic monopole and magnetic toroidal dipole.
Lastly, we summarize the results in Sec.~\ref{sec:summary}. 
In Appendix~\ref{app:CEF_dependence}, we show the CEF dependence of the magnetic anisotropy. 
In Appendix~\ref{app:ETD_scf}, we show the finite-temperature phase diagram when the exchange interaction for the ETD is considered.
In Appendix~\ref{app:E16sl}, we briefly show the result under the electric hexadecapole moment, which is another candidate hosting the ferroaxial moment in some crystals.

\section{Magnetic anisotropy under ferroaxial moment}
\label{sec:anisotropy}
We discuss the role of the ETD moment on magnetic anisotropy. 
In Sec.~\ref{subsec:ferroaxial_moment}, we introduce the ETD moment, which corresponds to a ferroaxial moment. 
We also show when the ETD degree of freedom is activated in the Hilbert space. 
In Sec.~\ref{subsec:single-site_model}, we introduce a single-site five $d$-orbital model. 
Then, we perform the second-order perturbation theory by focusing on the role of the ETD moment in Sec.~\ref{subsec:anisotropy}.
Finally, we show the magnetic anisotropy within the mean-field calculations in Sec.~\ref{subsec:1site_result}. 

\subsection{Microscopic description of ferroaxial moment}
\label{subsec:ferroaxial_moment}
\begin{figure}[th]
  \begin{center}
  \includegraphics[width=\hsize]{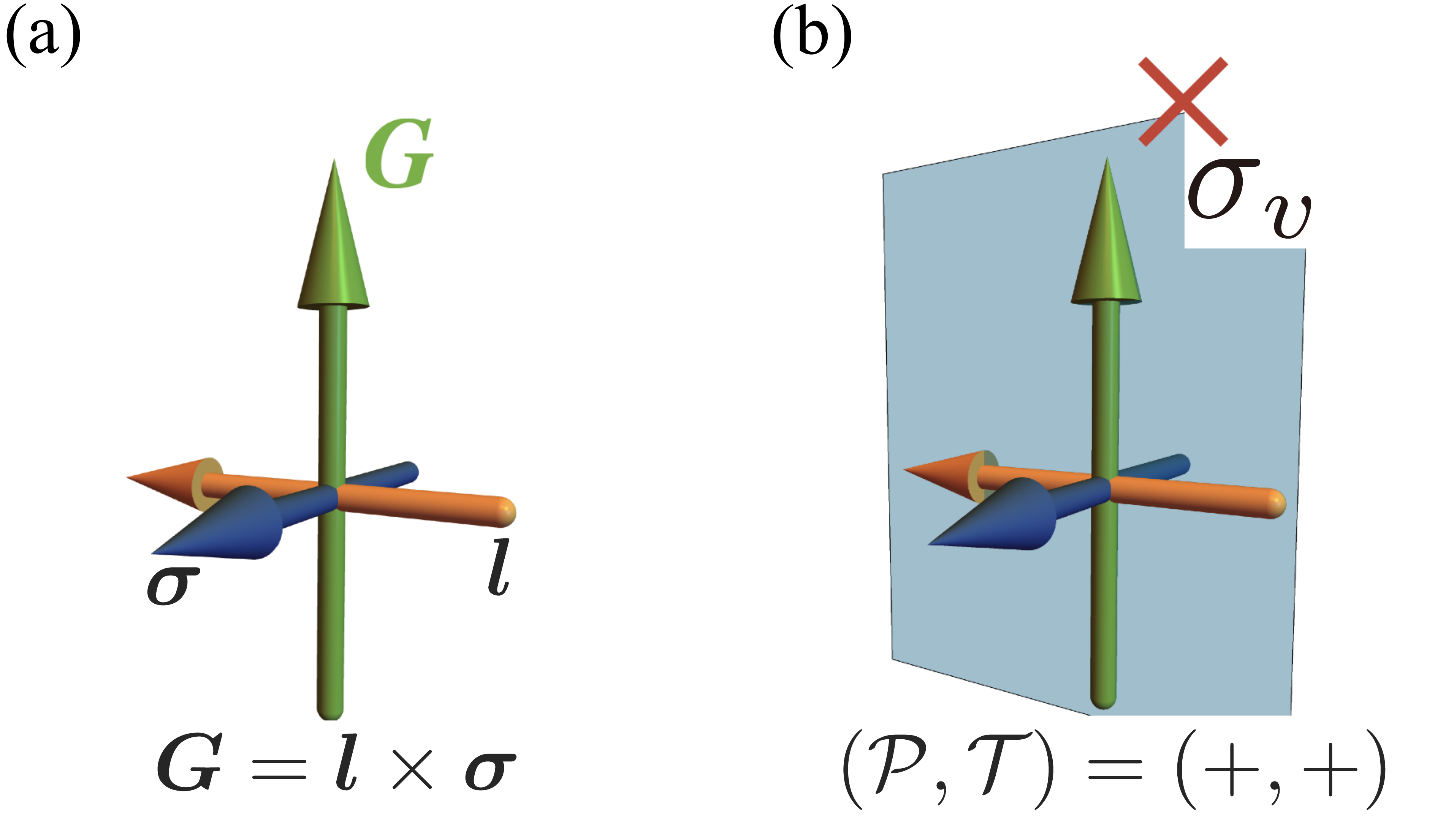} 
  \caption{
  \label{Fig:Gz}
  (a) The representation of the atomic-scale ETD moment $\bm{G} = \bm{l} \times \bm{\sigma}$, which is denoted by the green arrow. 
  The orange (blue) arrow represents the orbital (spin) angular momentum $\bm{l} (\bm{\sigma})$.
  (b) The symmetry of the atomic ETD moment; the mirror symmetry parallel to the ETD moment, $\sigma_v$, is lost. 
  Meanwhile, time-reversal ($\mathcal{T}$) and spatial inversion ($\mathcal{P}$) symmetries are retained.
  }
  \end{center}
\end{figure}
The ferroaxial moment can appear when the mirror symmetry parallel to the moment direction is lost; the symmetry breakings in terms of $\mathcal{P}$ and $\mathcal{T}$ are not necessary.
It is microscopically characterized by a ferroic alignment of a $\mathcal{T}$-even axial vector, which is referred to as the ETD $\bm{G}$. 
Based on the multipole description~\cite{hayami2018prb_Classification_of_atomic-scale_multipoles, kusunose2020complete}, the atomic-scale $\bm{G}$ operator is represented as the outer product of the spin operator $\bm{s} = \bm{\sigma}/2$ and orbital angular momentum operator $\bm{l}$ as follows:
\begin{align}
  \label{eq:atomic_G}
  \bm{G} = \bm{l} \times \bm{\sigma},
\end{align}
where the schematic picture of $\bm{G}$ is shown in Fig.~\ref{Fig:Gz}(a).
It is noted that $\bm{G}$ can appear when the expectation values of $\bm{l}$ and $\bm{\sigma}$ are zero. 
Since both $\bm{l}$ and $\bm{\sigma}$ are axial vectors, the mirror symmetry parallel to $\bm{G}$ is broken, as shown in Fig.~\ref{Fig:Gz}(b). 
$\bm{G}$ can be activated in the Hilbert space possessing these two operators, such as the $p$, $d$, and $f$ orbitals.

Among them, we consider five $d$ orbitals in the low-energy Hilbert space in the following analysis.
In this situation, $\bm{G}$ is defined in the off-diagonal space between two orbitals with different total angular momenta $J=3/2$ and $J=5/2$. 

\subsection{Single-site $d$-orbital model}
\label{subsec:single-site_model}
In order to investigate the role of the ETD moment on the magnetic anisotropy, we consider a single-site five $d$-orbital model with $(d_u, d_v, d_{yz}, d_{zx}, d_{xy})$ for $u=3z^2-r^2$ and $v=x^2-y^2$, which is given by
\begin{eqnarray}
  \label{eq:Ham}
  \mathcal{H}&=&\mathcal{H}^{\rm loc}+\mathcal{H}^{\rm ex}, \\
  \label{eq:1site_Hamiltonian}
  \mathcal{H}^{\rm loc} &=& \mathcal{H}^{\rm CEF} + 
  \lambda \bm{l} \cdot \bm{s} -h_G G_z, \\
  \label{eq:1site_EX}
  \mathcal{H}^{\rm ex} &=& -J_0 (s_{x}^2 + s_{y}^2), 
\end{eqnarray}
where the first term in Eq.~(\ref{eq:Ham}) represents the one-body Hamiltonian, while the second term represents the two-body Hamiltonian.
In $\mathcal{H}^{\rm loc}$, $\mathcal{H}_{\rm CEF}$ is the crystalline electric filed (CEF) Hamiltonian. 
We consider the five CEF parameters by supposing the $D_{\rm 2h}$ symmetry: $\Delta_1 = 0.400$, $\Delta_2 \simeq
1.448$, $\Delta_3 = 2.200$, $\Delta_4 \simeq 2.552$, $\alpha \simeq 0.572$, and $\beta = \sqrt{1-\alpha^2}$, as schematically shown in Fig.~\ref{Fig:model}(a); $\Delta_1$--$\Delta_4$ denote the atomic energy levels for $d_{xy}$, $-\alpha d_u + \beta d_v$, $d_{zx}$, and $\beta d_u + \alpha d_v$ orbitals measured from that for the $d_{yz}$ orbital, where $\alpha$ and $\beta$ stand for the numerical coefficients; see Appendix~\ref{app:CEF_dependence} for the detailed definition of $\mathcal{H}^{\rm CEF}$. 
We suppose that the ground-state energy level is the $d_{yz}$ orbital. 
We take the principal axis along the $z$ direction, as shown in Fig.~\ref{Fig:model}(b). 
The second term in Eq.~(\ref{eq:1site_Hamiltonian}) represents the atomic SOC.
The third term in Eq.~(\ref{eq:1site_Hamiltonian}) represents the molecular field that arises from the ETD moment, which lowers the symmetry from $D_{\rm 2h}$ to $C_{\rm 2h}$. 

\subsection{Perturbation analysis}
\label{subsec:anisotropy}
\begin{figure}[t]
  \begin{center}
  \includegraphics[width=\hsize]{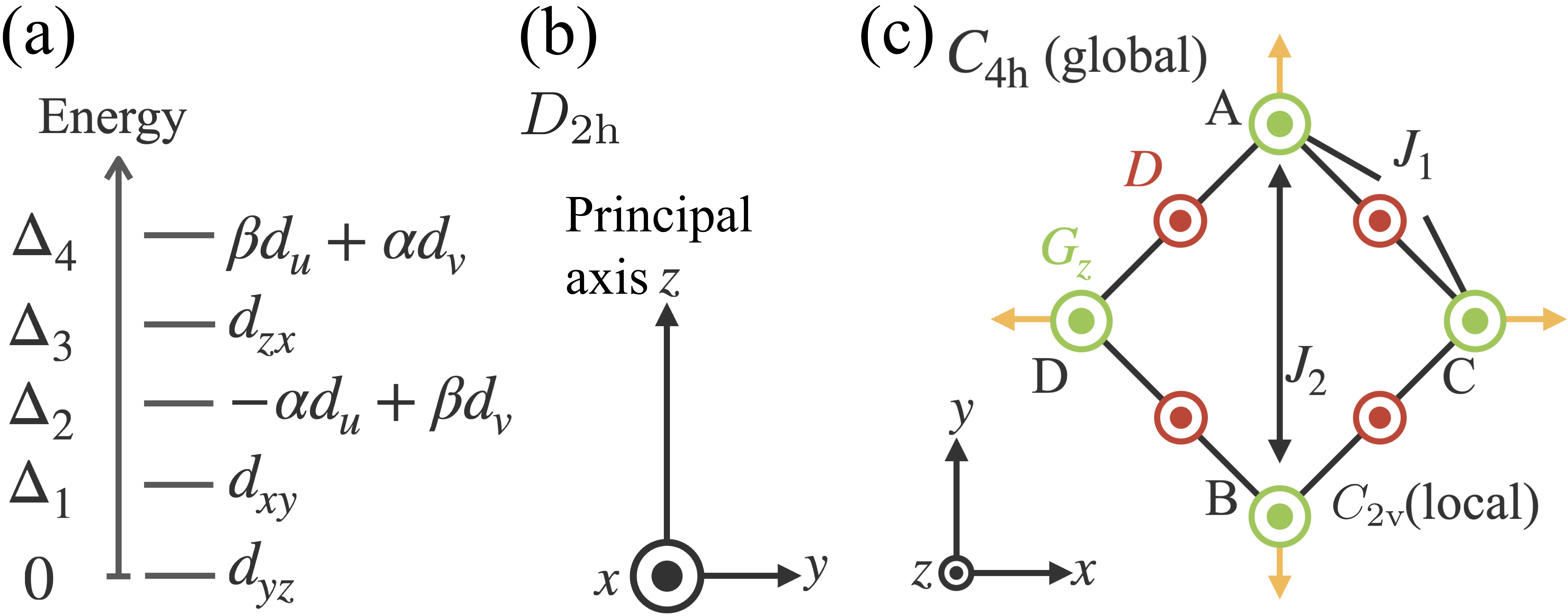} 
  \caption{
  \label{Fig:model}
  (a) A schematic picture of the CEF levels for the single-site five $d$-orbital model under the $D_{\rm 2h}$ symmetry. 
  (b) The coordinate axes for the single site, where the $z$ axis is taken as the principal axis. 
  (c) Four-site tetragonal cluster under the $C_{\rm 4h}$ symmetry, where the site symmetry is $C_{\rm 2v}$. 
  $J_1$ and $J_2$ represent the exchange interactions between the nearest-neighbor and next-nearest-neighbor sites, respectively. 
  Yellow arrows represent the electric polarization $\bm{P}$, which originates from the potential gradient at each site. 
  Green (red) circles represent ETD moments (DM vectors) along the $z$ direction. 
  }
  \end{center}
\end{figure}

We examine the magnetic anisotropy arising from the ETD moment by performing the perturbation analysis. 
For that purpose, we analyze $\mathcal{H}^{\rm loc}$ in Eq.~(\ref{eq:1site_Hamiltonian}) by ignoring $\mathcal{H}^{\rm ex}$ in Eq.~(\ref{eq:1site_EX}). 
Within the second-order perturbation in terms of $\lambda$ and $h_G$, an effective spin Hamiltonian is derived as
\begin{align}
  \mathcal{H}_{\rm s}  &= -\sum_{\mu, \nu = x,y}\Lambda_{\mu\nu} s_{\mu}s_\nu ,\\  
  \label{eq:Lambda_tot}
    \bm{\Lambda} 
    &= 
    \begin{bmatrix}
    4 h^2_{G} \Lambda'_{yy} + \lambda^2 \Lambda'_{xx} &-2h_{G}  \lambda (\Lambda'_{xx} -\Lambda'_{yy}) \\
    -2h_{G} \lambda (\Lambda'_{xx} -\Lambda'_{yy}) & 4 h^2_{G} \Lambda'_{xx} + \lambda^2 \Lambda'_{yy}
    \end{bmatrix},
\end{align}
where 
\begin{align}
  \label{eq:Lambda}
  \Lambda'_{\mu\nu} &=  \sum_e\frac{\langle g|l_\mu|e \rangle \langle e|l_\nu|g \rangle }{E_e-E_g}.
\end{align}
$ \Lambda'_{\mu\nu}$ includes the contribution from the CEF, where $\ket{g}(\ket{e})$ is the ground state (excited state) and $E_g (E_e)$ is the ground-state (excited-state) energy. 
We here omit the $z$ component of $\Lambda_{\mu\nu}$, since the effect of the ETD does not appear in $\Lambda_{z\nu}$ and $\Lambda_{\mu z }$.

There are three important observations in Eq.~(\ref{eq:Lambda_tot}).
One is the emergence of the off-diagonal $xy$ component in $\Lambda_{\mu\nu}$. 
Thus, the ferroaxial moment induced by $h_{G}$ tends to tilt the spin moment from the crystal axis. 
In addition, it is noteworthy that the SOC $\lambda$ is necessary to induce the off-diagonal component. 
The second is the importance of the low-symmetric CEF to induce $\Lambda_{xy}$, since it is proportional to $\Lambda'_{xx} -\Lambda'_{yy}$ for $\Lambda'_{xx} \neq \Lambda'_{yy}$. 
In other words, the inequivalence between the $x$ and $y$ directions is significant. 
This is why we consider the orthorhombic CEF Hamiltonian under the $D_{\rm 2h}$ symmetry in $\mathcal{H}^{\rm CEF}$; $\Lambda_{xy}=0$ when the tetragonal and hexagonal CEFs are considered. 
The last is the opposite tendency in the diagonal component of $\Lambda_{\mu\nu}$ for $\mu=\nu$ between the ETD moment and the SOC; $h_G$ ($\lambda$) tends to favor the $x$ ($y$) direction for $\Lambda'_{xx} < \Lambda'_{yy}$.  

More specifically, one obtains the tilt angle $\theta$ from the $x$ axis by diagonalizing $\bm{\Lambda}$, which is given by $\theta = \arctan[\lambda/(2h_G)]$ for $\Lambda'_{xx} < \Lambda'_{yy}$ or $\arctan(- 2h_G/\lambda) $ for $\Lambda'_{xx} > \Lambda'_{yy}$. 
In the case of the CEF parameters in Fig.~\ref{Fig:model}(a), $\Lambda'_{xx} < \Lambda'_{yy}$.
We discuss the magnetic anisotropy for the other CEF levels in Appendix~\ref{app:CEF_dependence}.

\subsection{Mean-field calculations}
\label{subsec:1site_result}
\begin{figure}[t]
  \begin{center}
  \includegraphics[width=\linewidth]{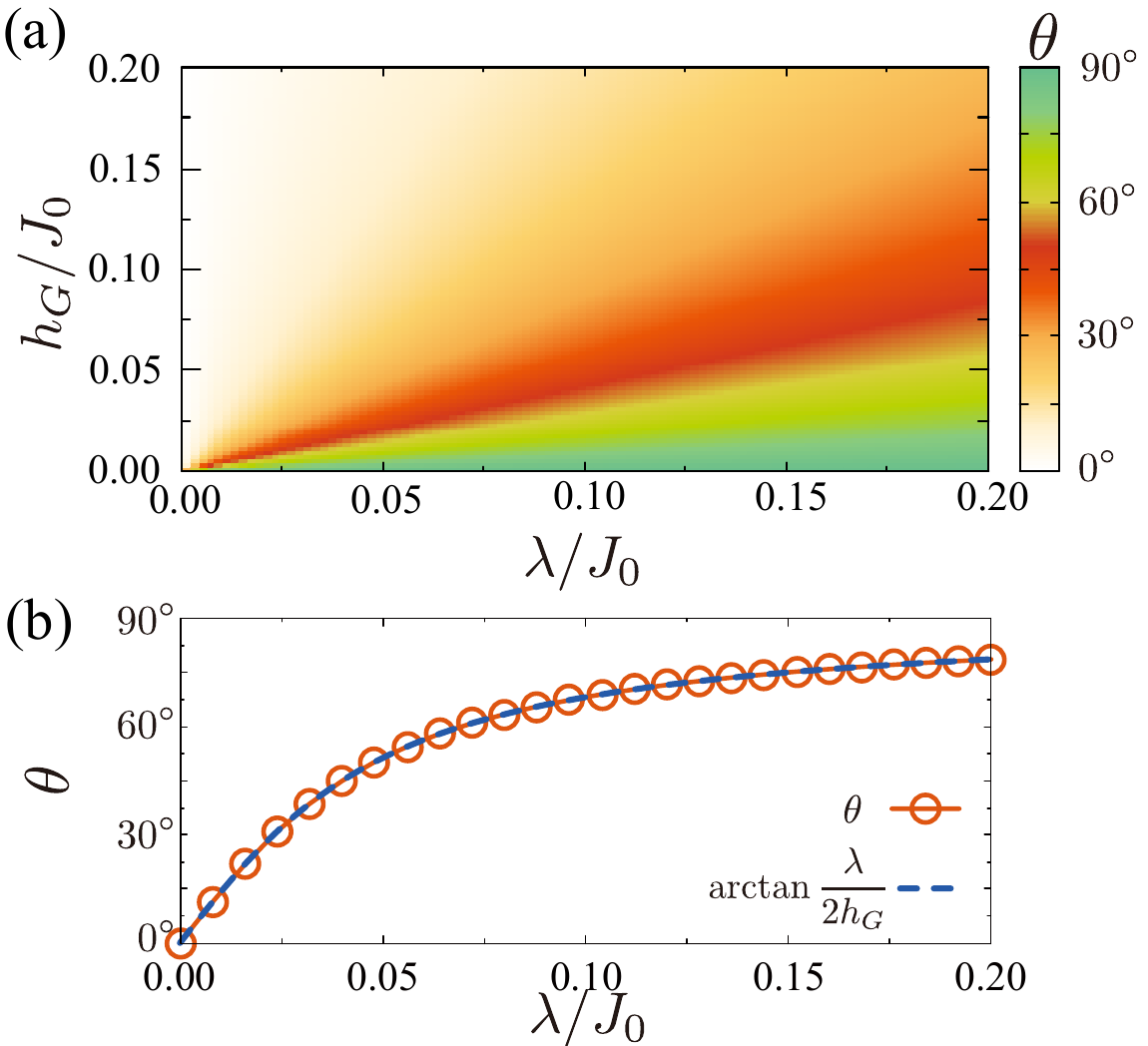} 
  \caption{
  \label{Fig:1site_theta}
  (a) Contour plot of the tilt angle of a spin moment $\theta$ in the plane of $\lambda$ and $h_G$, which is obtained by the mean-field calculations. 
  (b) $\lambda$ dependence of the numerical result (red circle) and the perturbation result (blue dashed line) at $h_G=0.02$. In both panels, the data is calculated by changing $\lambda$ and/or $h_G$ with the interval of $\Delta_\lambda/J_0 = \Delta_{h_G}/J_0 = 0.002$.
  }
  \end{center}
\end{figure}

We numerically evaluate the tilt angle $\theta$ in the presence of the two-body Hamiltonian $\mathcal{H}^{\rm ex}$. 
We apply the mean-field approximation for $\mathcal{H}^{\rm ex}$ as 
\begin{eqnarray}
  \label{eq: mean-field_approx}
  \mathcal{H}_{\rm MF}^{\rm ex} &=& -
  J_0
  (\braket{s_{x}}s_{x} + \braket{s_{y}}s_{y}) + \rm (const.),
\end{eqnarray}
where $\braket{\cdots}$ represents the statistical average in $d^1$ configuration.
We set $J_0$ to the energy unit of the single-site model ($J_0=1$). 

Figure~\ref{Fig:1site_theta}(a) shows the angle $\theta$ of the spin moment measured from the $x$ axis by changing $\lambda$ and $h_G$ at temperature $T/J_0=0.1$.
When either $\lambda$ or $h_G$ becomes zero, the spin aligns in the crystal axis. 
For $\lambda/J_0=0$, the spin moment aligns in the $x$ direction, while it aligns in the $y$ direction for $h_G/J_0=0$. 
This feature is consistent with the perturbation analysis, where $\Lambda'_{xx} < \Lambda'_{yy}$ is satisfied. 

Meanwhile, the spin tilts from the crystal axis when both $\lambda$ and $h_G$ are considered. 
One finds good agreement between numerical and perturbation results; see the $\lambda$ dependence of both results in the case of $h_G/J_0=0.02$ in Fig.~\ref{Fig:1site_theta}(b).
In addition, the maximum tilt angle of $\theta=45^\circ$ is realized for $\lambda \simeq 2h_G$, which is also consistent with the perturbation result in Eq.~(\ref{eq:Lambda_tot}); the feature holds when $\lambda$ increases. 
These results indicate that the interplay between $\lambda$ and $h_G$ plays an important role in inducing the magnetic anisotropy characteristic of the ETD moment even beyond the perturbation regime.

\section{magnetic instability in a cluster model}
\label{sec:instability}
\subsection{Four-site cluster model}
\label{subsec:4-site_model}
\begin{figure}[t]
  \begin{center}
  \includegraphics[width=\linewidth]{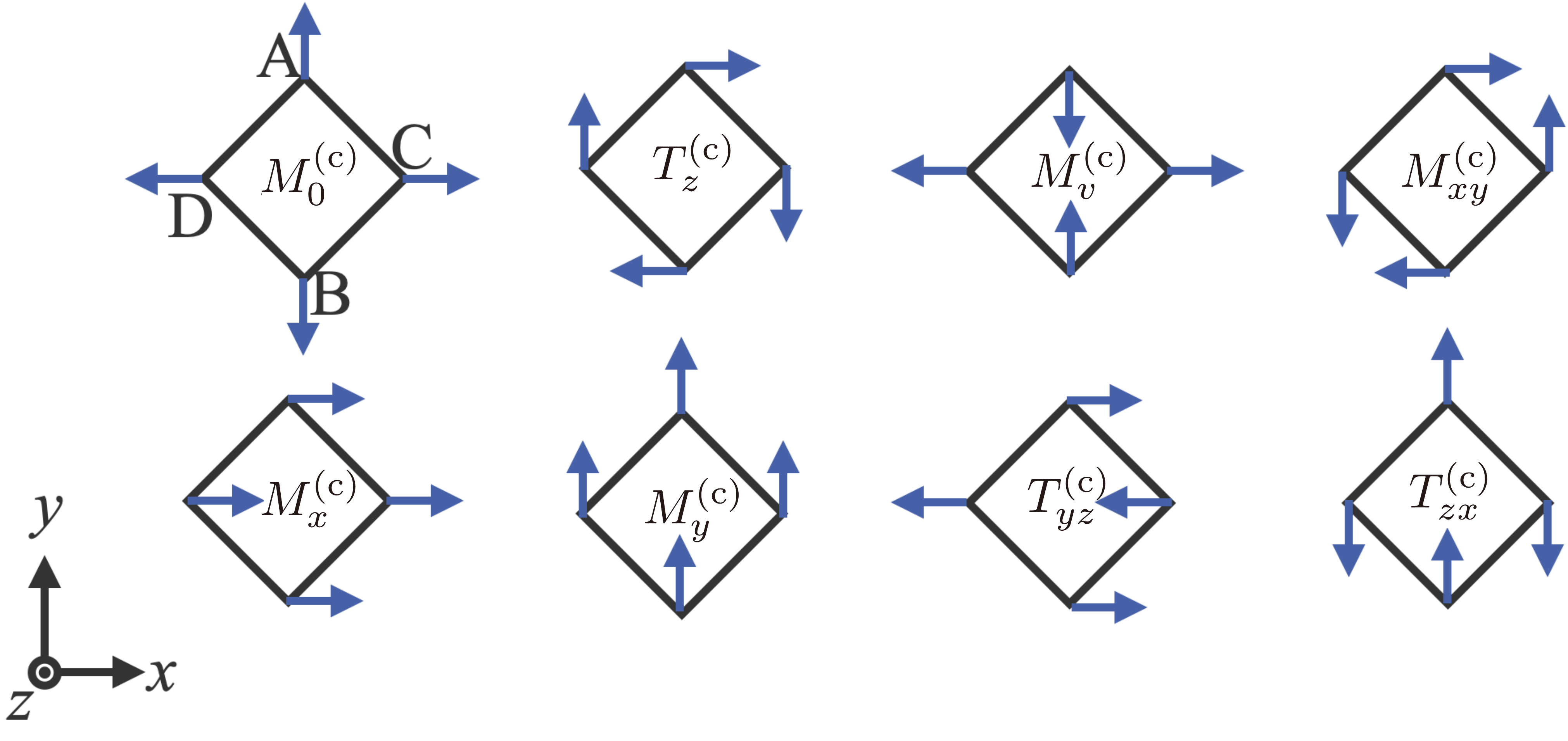} 
  \caption{
  \label{Fig:8_spin_arrangements}
  Eight independent spin configurations in the four-site cluster: magnetic monopole ($M^{(\rm c)}_0$), magnetic dipole ($M^{(\rm c)}_x, M^{(\rm c)}_y$), magnetic toroidal dipole ($T^{(\rm c)}_z$), magnetic quadrupole ($M^{(\rm c)}_{xy}, M^{(\rm c)}_v$), and magnetic toroidal quadrupole ($T^{(\rm c)}_{yz}, T^{(\rm c)}_{zx}$).  
  Blue arrows represent the spin moments at each site.
  }
  \end{center}
\end{figure}

Next, let us consider the magnetic instability under the ETD moment in a cluster system by extending the analysis to the single-site system.
We consider a four-site tetragonal cluster under the $C_{\rm 4h}$ symmetry, as shown in Fig.~\ref{Fig:model}(c). 
We suppose that the site symmetry is $C_{\rm 2v}$ so that the off-diagonal component in $\Lambda_{\mu\nu}$ in Eq.~(\ref{eq:Lambda_tot}) becomes nonzero. 
We use the local Hamiltonian $\mathcal{H}^{\rm loc}$ in Sec.~\ref{subsec:single-site_model} by adding the sublattice index $i=$ A--D. 
We take the same CEF parameters in Sec.~\ref{sec:anisotropy}, although $\alpha$ take opposite sign and $d_{yz} \leftrightarrow d_{zx}$ 
between (A, B) and (C, D) sublattices because the principal axis for the A and B sublattices is different from that for the C and D sublattices by 90$^{\circ}$.

For the exchange interaction, we consider the following Hamiltonian, which is given by 
\begin{eqnarray}
  \mathcal{H}'^{\rm ex} &=&  - J_1
  \sum^{{\rm n.n}}_{ \langle i,j \rangle} 
  (s^i_{x}s^j_{x} + s^i_{y}s^j_{y}) \notag\\
  & & -J_2
  \sum^{\rm{n.n.n}}_{ \langle i,j \rangle} 
  (s^i_{x}s^j_{x} + s^i_{y}s^j_{y}),
\end{eqnarray}
where $J_1$ and $J_2$ correspond to the coupling constants for the nearest-neighbor (n.n.) and next-nearest-neighbor (n.n.n) sites, respectively. 
We here consider the situation where magnetic ordering with the in-plane spin modulations occurs rather than the out-of-plane ones.
In addition, we consider the DM interaction $\bm{D}=(D_x,D_y,D_z)$. 
From the symmetry viewpoint, only the $z$ component between the nearest-neighbor sites becomes finite, as shown in Fig.~\ref{Fig:model}(c). 
The DM Hamiltonian is given by
\begin{eqnarray}
\mathcal{H}^{\rm DM} = 
-\sum^{{\rm n.n}}_{ \langle i,j \rangle} D_z^{ij}
(\boldsymbol{s}^i\times\boldsymbol{s}^j)_z,
\end{eqnarray}
where $D_z^{ij}=-D_z^{ji}\equiv D$.
By adopting the mean-field approximation, 
$\mathcal{H}'^{\rm ex}$ and $\mathcal{H}^{\rm DM}$ are represented as 
\begin{eqnarray}
  \label{eq:4site_mean-field}
  \mathcal{H}^{'\rm ex}_{\rm MF} &\simeq&  
  -J_1 \sum_i^{\rm A,B,C,D}\sum_j^{\rm n.n}
  \Big(\Braket{s^{i}_{x}} s^{j}_{x}+\Braket{s^{i}_{y}} s^{j}_{y}\Big) \notag\\
  &&- J_2 \sum_i^{\rm A,B,C,D}\sum_j^{\rm n.n.n}
  \Big(\Braket{s^{i}_{x}} s^{j}_{x}+\Braket{s^{i}_{y}} s^{j}_{y}\Big),\\
  \mathcal{H}_{\rm MF}^{\rm DM} &\simeq& 
  -D \sum_i^{\rm A,B,C,D}\sum_j^{\rm n.n}[\braket{\boldsymbol{s}^i}\times\boldsymbol{s}^j]_z,
\end{eqnarray}
where we omit the constant term for notational simplicity. 
We set $J_1$ to the energy unit of the four-site cluster model ($J_1=1$).
Although we treat the effect of the ETD as the one-body mean field $h_G$ for simplicity, a qualitatively similar result can be obtained even when the ETD moment is induced through the two-body exchange interaction, as discussed in Appendix~\ref{app:ETD_scf}.

\subsection{Spin configurations}
We consider the magnetic instability in the four-site cluster model within the mean-field approximation. 
Since we suppose the in-plane magnetic anisotropy, the four-sublattice magnetic structures are expressed as a linear combination of eight independent spin configurations.
Based on the cluster multipole theory~\cite{Suzuki_PhysRevB.99.174407}, they are classified into magnetic and magnetic toroidal multipoles:
magnetic monopole ($M^{(\rm c)}_0$), magnetic dipole ($M^{(\rm c)}_x, M^{(\rm c)}_y$), magnetic toroidal dipole ($T^{(\rm c)}_z$), 
magnetic quadrupole ($M^{(\rm c)}_{xy}, M^{(\rm c)}_v$), and magnetic toroidal quadrupole ($T^{(\rm c)}_{yz}, T^{(\rm c)}_{zx}$). 
Specifically, their spin configurations denoted as $(\sigma^{\rm A}_x,\sigma^{\rm A}_y,\sigma^{\rm B}_x,\sigma^{\rm B}_y,\sigma^{\rm C}_x,\sigma^{\rm C}_y,\sigma^{\rm D}_x,\sigma^{\rm D}_y)$ are given by
\begin{align}
  M^{(\rm c)}_0 &=(0,1, 0,-1, 1,0, -1,0),\\ 
  M^{(\rm c)}_x &= (1,0, 1,0, 1,0, 1,0),\\ 
  M^{(\rm c)}_y &= (0,1, 0,1, 0,1, 0,1),\\ 
  T^{(\rm c)}_z &= (1,0, -1,0, 0,-1, 0,1),\\ 
  M^{(\rm c)}_{xy} &=(1,0, -1,0, 0,1, 0,-1),\\
  M^{(\rm c)}_v &= (0,-1, 0,1, 1,0, -1,0),\\
  T^{(\rm c)}_{yz} &=(1,0, 1,0, -1,0, -1,0),\\
  T^{(\rm c)}_{zx} &=(0,1, 0,1, 0,-1, 0,-1).
\end{align}

The lowest-energy spin configuration depends on the magnetic interactions $(J_1, J_2, D)$ as well as $\mathcal{H}^{\rm loc}$. 
When $D/J_1=0$, $|J_1| < |J_2|$, 
$J_2 < 0$, and $\mathcal{H}^{\rm loc}$ is negligible, the energy by the exchange interactions becomes the lowest for any of $M^{(\rm c)}_0, T^{(\rm c)}_z, M^{(\rm c)}_v$, and $M^{(\rm c)}_{xy}$. 
In this situation, by introducing $D>0$, the energy for $M^{(\rm c)}_0$ and $T^{(\rm c)}_z$ is smaller than that for $M^{(\rm c)}_v$ and $M^{(\rm c)}_{xy}$. 
With this tendency in mind, we take $|J_1| \sim |J_2|$, $J_2 < 0$, and $D/J_1=0.05>0$. 
Although the energy for $M^{(\rm c)}_0$ and $T^{(\rm c)}_z$ is degenerate with each other, it splits
by taking into account the effect of the $\mathcal{H}^{\rm loc}$.

\subsection{Magnetic phase diagram}
\label{subsec:magnetic_phase_diagram}
\begin{figure}[th]
  \centering
  \includegraphics[width=\linewidth]{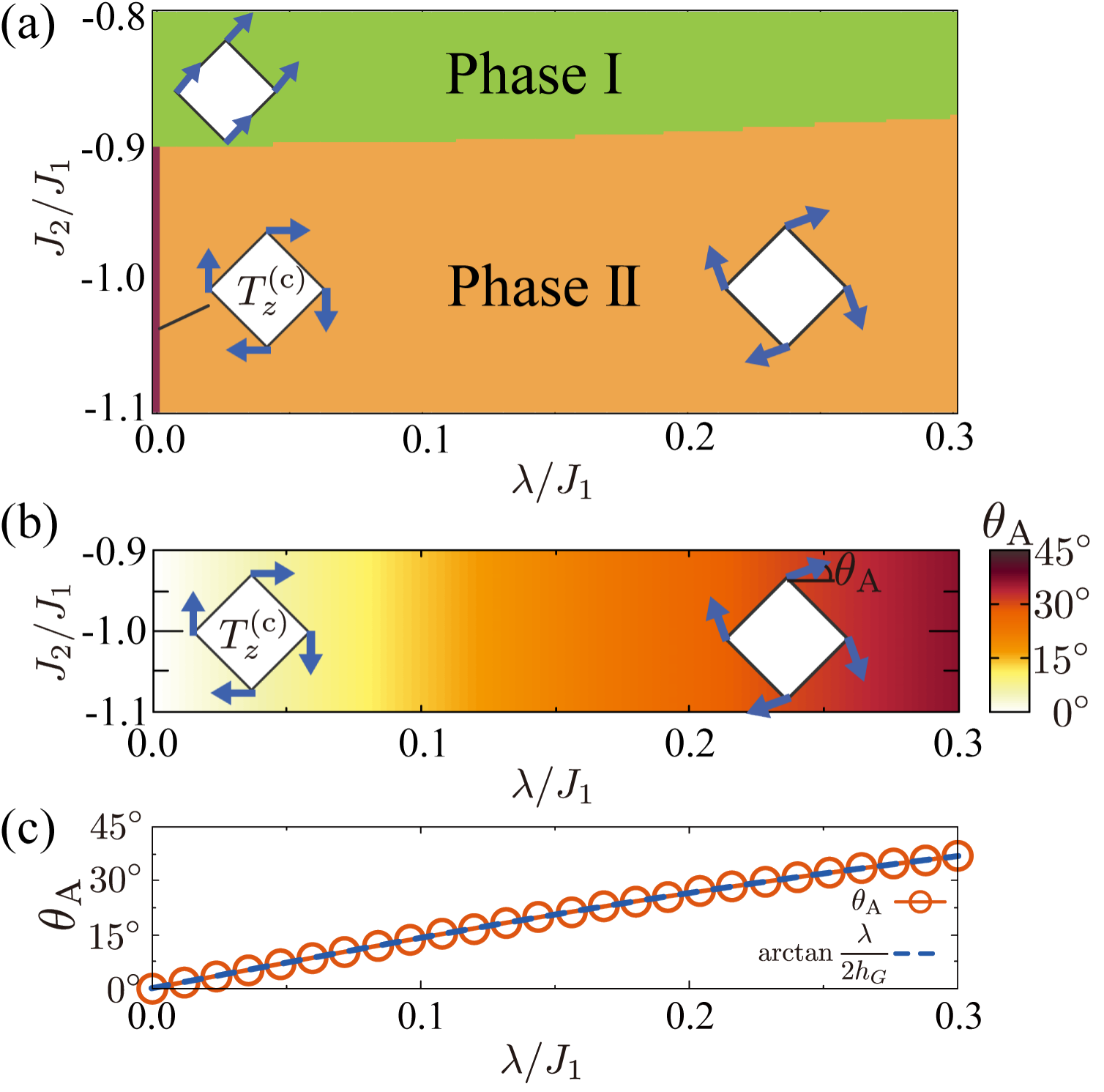} 
  \caption{
  \label{Fig:4site_phase}
  (a) $\lambda$--$J_2$ phase diagram obtained at $h_G/J_1 = 0.2$  
  and $D/J_1 = 0.05$. 
  The data is calculated by changing $\lambda$ and $h_G$ with the interval of $\Delta_\lambda/J_1 = \Delta_{J_2}/J_1 = 0.003$. Phase\Rnum{1} (\!\!\Rnum{2}) represents the spin configuration consisting of the linear combination of $M^{\rm (c)}_x, M^{\rm (c)}_y, T^{\rm (c)}_{yz}$, and $T^{\rm (c)}_{zx}$ ($M^{\rm (c)}_0$ and $T^{\rm (c)}_z$) in Fig.~\ref{Fig:8_spin_arrangements}. 
  At $\lambda/J_1 = 0$ for $J_2/J_1< -0.9$, the $T^{\rm (c)}_z$ state is stabilized. 
  (b) Contour plot of the tilt angle of spin moments at A sublattice $\theta_{\rm A}$ for $-1.1 \leq J_2 /J_1 \leq -0.9$. 
  (c) $\lambda$ dependence of the numerical result (red circle) and the perturbation result (blue dashed line) in terms of $\theta_{\rm A}$ at $J_2/J_1 = -1.1$.
  }
\end{figure}

We perform the self-consistent mean-field calculations for the four-site cluster model by setting $h_G/J_1=0.2$ and the temperature as $T/J_1=0.1$. Figure~\ref{Fig:4site_phase}(a) shows the magnetic phase diagram by changing $\lambda$ and $J_2$.
There are mainly two phases in the phase diagram: Phase I and Phase II. 

The spin configuration of Phase I is expressed as the linear combination of $M^{\rm (c)}_x, M^{\rm (c)}_y, T^{\rm (c)}_{yz}$, and $T^{\rm (c)}_{zx}$; the spin moments tilt from the crystal axis owing to the magnetic anisotropy arising from $h_G$.
Although Phase I is almost characterized by the ferromagnetic spin configuration, i.e., $M^{\rm (c)}_x$ and $M^{\rm (c)}_y$, it includes the small contribution from $T^{\rm (c)}_{yz}$ and $T^{\rm (c)}_{zx}$. 
This is because the principal axis for the A and B sublattices is different from that for the C and D sublattices by 90$^{\circ}$, which means that $x$ and $y$ axes in the global coordinate [Fig.~\ref{Fig:model}(c)] are inequivalent for the (A, B) and (C, D) sublattices, and hence, the spin lengths between them are different from each other when the moments lie in the $xy$ plane in a uniform way.

Meanwhile, Phase II is characterized by the spin configuration to possess the fourfold rotational symmetry in order to gain the energy by the CEF. 
For $J_2/J_1 < -0.9$ and $\lambda/J_1=0$, the spin configuration in Phase II corresponds to $T^{\rm (c)}_z$. 
By introducing $\lambda$, the spins at four sublattices tilt in the same manner so as to keep the fourfold rotational symmetry, which indicates that the spin configuration is expressed as the linear combination of $M^{\rm(c)}_0$ and $T^{\rm(c)}_z$, as schematically shown in the inset of Fig.~\ref{Fig:4site_phase}(a).

We discuss the effect of $\mathcal{H}^{\rm loc}$ including the ETD molecular field $h_G$ and the SOC $\lambda$ for $J_2/J_1 <-0.9$.
In the case of $\mathcal{H}^{\rm loc}=0$, the phase boundary between Phase I and Phase II is given by $J_2/J_1=-0.95$. 
Thus, $\mathcal{H}^{\rm loc}$ tends to favor the region for Phase II. 
This is understood from the fact that $\mathcal{H}^{\rm loc}$, which becomes the origin of the magnetic anisotropy, favors the spin configuration satisfying the fourfold rotational symmetry that the four-site cluster possesses in order to gain the energy by the magnetic anisotropy. 
Such a tendency holds for nonzero $\lambda$, which enhances the magnetic anisotropy; the phase boundary moves upward by increasing $\lambda$. 
Thus, both $h_{G}$ and $\lambda$ tend to favor the vortex spin configuration retaining the fourfold rotational symmetry compared to the uniform one breaking the fourfold rotational symmetry.

Figure~\ref{Fig:4site_phase}(b) shows the tilt angle of spin moments at A sublattice $\theta_{\rm A}$ in Phase\Rnum{2} for $-1.1 \leq J_2/ J_1 \leq -0.9$.
The behavior is similar to that in the single-site model; the tilt angle increases as $\lambda$ increases.
Furthermore, we confirmed that such behavior is understood from the perturbation calculations, as shown in Fig.~\ref{Fig:4site_phase}(c); both data are well consistent.

Finally, we discuss the relationship between the ETD moment and vortex magnetic structures from the symmetry viewpoint.
Since $M_0$ corresponds to a time-reversal-odd axial scalar and $T_z$ corresponds to a time-reversal-odd polar vector, their product $M_0 T_z$ corresponds to a time-reversal-even axial vector, i.e., the ETD $G_z$~\cite{hayami2022prb_spin_texture}.
In this sense, the appearance of Phase II, which is expressed as the linear combination of $M_0$ and $T_z$ in the presence of $G_z$, is natural. 
In a similar context, it was shown that the skyrmion crystal accompanying both $M_0$ (N\'eel type) and $T_z$ (Bloch type) is realized by considering the magnetic anisotropy that originates from the mirror symmetry breaking~\cite{Hayami_PhysRevB.105.104428}. 
Since $M_0$ and $T_z$ lead to similar but different physical phenomena, the coexisting state can give rise to further intriguing cross-correlation responses and transports. 
The electric axial moment, $G_z$, plays an important role in inducing such an effective coupling of $M_0$ and $T_z$.

\section{Summary}
\label{sec:summary}
To summarize, we have investigated the magnetic single-ion anisotropy and its associated magnetic instability driven by the ETD moments based on the perturbation and mean-field calculations for the five $d$-orbital models in the single-site and four-site cluster.
We show that the synergy between the molecular field arising from the ETD moment and the SOC is essential to induce the in-plane magnetic anisotropy so that the spin tilts from the crystal axis. 
We also show that the tendency to tilt the spin moments is enhanced when the ETD molecular field and SOC are comparable to each other. 
We discuss that the ferroaxial system might become a prototype to realize the vortex spin texture with both the nature of the magnetic monopole and magnetic toroidal dipole.
One of the candidate materials is CaMn$_7$O$_{12}$, where a in-plane spin vortex structure was identified in experiments~\cite{johnson2012prl_CaMn7O12}.

In addition, the present tendency in terms of magnetic anisotropy is also expected for other ferroaxial materials. 
Since the ferroaxial moment can appear in crystallographic point groups without mirror symmetry parallel to the electric axial moment, $C_{6\rm h}, C_6, C_{3\rm h}, C_{4\rm h}, C_4, S_4, C_{3\rm i}, C_3, C_{2\rm h}, C_2, C_{\rm s}, C_{\rm i}$, and $C_1$, the materials with these crystal structures can exhibit similar vortex spin configurations when the magnetic phase transition occurs. 

Let us comment on the difference between the ETD and other multipole degrees of freedom which might also correspond to the electric axial moments in some crystals.
Although the ETD and other multipoles are independent of each other in the rotational group, they often belong to the same irreducible representation according to the symmetry lowering. 
In the cases of the $C_{\rm 2h}$ and $C_{\rm 4h}$ symmetries discussed in Secs.~\ref{sec:anisotropy} and \ref{sec:instability}, respectively, the $xy(x^2-y^2)$ type of the electric hexadecapole also leads to similar mirror symmetry breaking. 
Thus, the electric hexadecapole is another candidate to describe the ferroaxial ordering. 
Indeed, we find that the electric hexadecapole also leads to the tilt of spin moments, although its behavior against the model parameters is different. 
We discuss the difference between the ETD and electric hexadecapole in Appendix~\ref{app:E16sl}.

\begin{acknowledgments}
This research was supported by JSPS KAKENHI Grants Numbers JP21H01037, JP22H04468, JP22H00101, JP22H01183, JP23K03288, JP23H04869, and by JST PRESTO (JPMJPR20L8) and JST CREST (JPMJCR23O4).
\end{acknowledgments}

\appendix
\section{CEF dependence of magnetic anisotropy}
\label{app:CEF_dependence}

\begin{figure}[th]
\centering
\includegraphics[width=\linewidth]{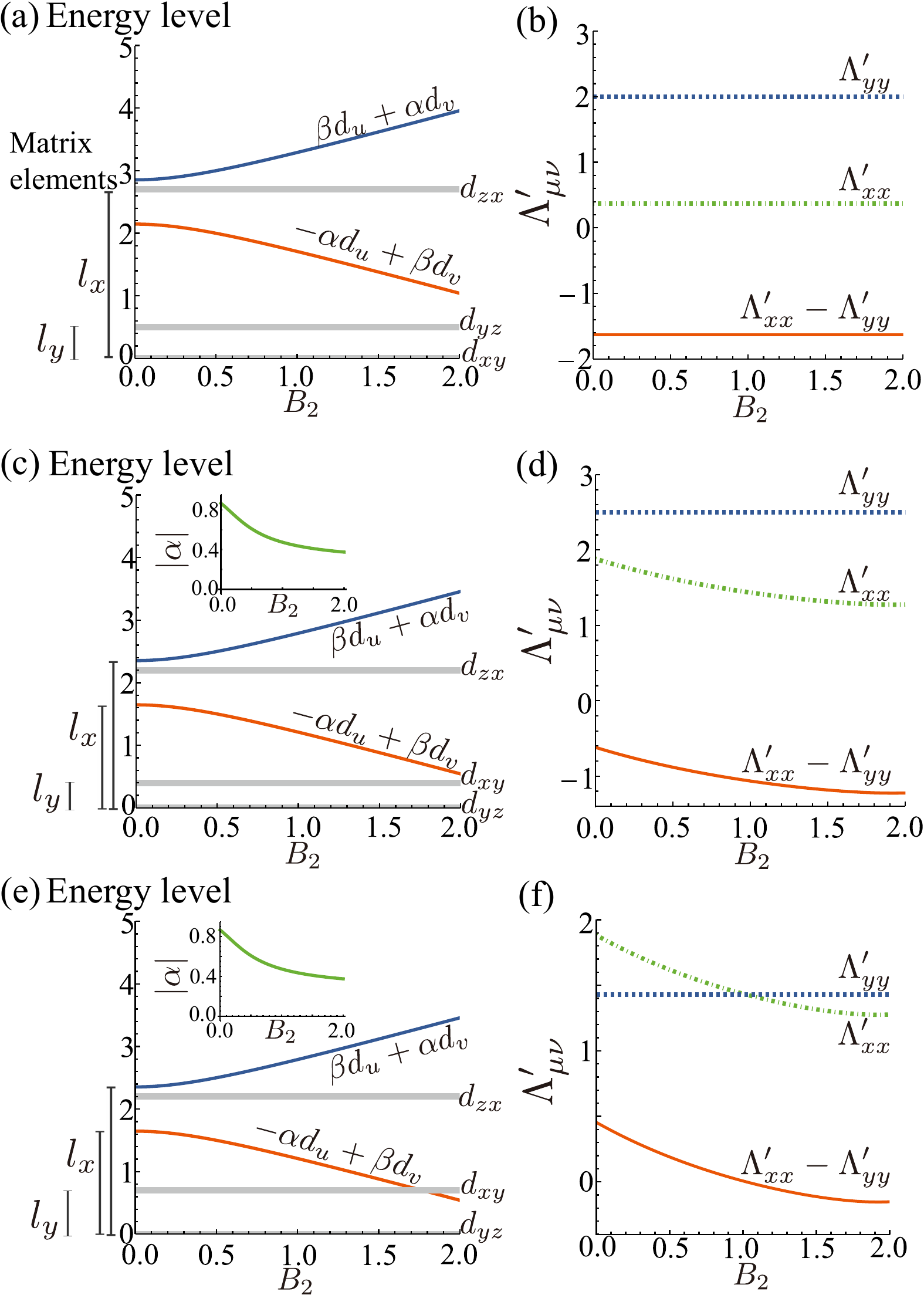} 
\caption{
$B_2$ dependence of (a), (c),
(e) energy levels and (b), (d), (f) $\Lambda'_{\mu\nu} (\mu, \nu = x,y)$ at $B_4 = -2.5$ in (a), (b), $B_4=-1.6$ in (c), (d), and $B_4=-1.3$ in (e), (f) for the single-site system. 
The red(blue) line represents the CEF energy level $-\alpha d_u + \beta d_v (\beta d_u + \alpha d_v)$. 
The insets of (c) and (e) show $B_2$ dependence of $|\alpha|$. 
The ground-state energy level is characterized by the $d_{xy}$ ($d_{yz}$) orbital for (a), (b) [(c), (d), (e), and (f)]. The other model parameters are chosen as $B_1 =0.5$, $B_3=-2$, and $B_5=0.2$. 
}
\label{Fig:CEF_dependence}
\end{figure}
In this Appendix, we show the relation between CEF parameters and the magnetic anisotropy. 
As shown in Eq.~(\ref{eq:Lambda}), the CEF energy levels and the matrix elements of orbital angular momentum $l_x, l_y$ affect the magnitude of anisotropy $\Lambda'_{xx}, \Lambda'_{yy}$. 
The matrix elements of $l_x$ and $l_y$ are given by 
\begin{eqnarray}
  l_x &=&
  \begin{pmatrix}
    0 & 0 & i\sqrt{3} & 0 & 0 \\
    0 & 0 & i & 0 & 0 \\
    -i\sqrt{3} & -i & 0 & 0 & 0\\
    0 & 0 & 0 & 0 & i\\
    0 & 0 & 0 & -i & 0\\
  \end{pmatrix},\\
  l_y &=&
  \begin{pmatrix}
    0 & 0 & 0 & -i\sqrt{3} & 0\\
    0 & 0 & 0 & i & 0\\
    0 & 0 & 0 & 0 & -i\\
    i\sqrt{3} & -i& 0 & 0 & 0\\
    0 & 0 & i & 0 & 0 \\
  \end{pmatrix},
\end{eqnarray}
where the basis is given by five $d$ orbitals: $\ket{d_u}, \ket{d_v}, \ket{d_{yz}}, \ket{d_{zx}}, \ket{d_{xy}}$. 
$l_x (l_y)$ has the matrix elements between $d_u$ and $d_{yz}$, $d_v$ and $d_{yz}$, and $d_{zx}$ and $d_{xy}$ ($d_u$ and $d_{zx}$, $d_v$ and $d_{zx}$, and $d_{yz}$ and $d_{xy}$).  
To investigate the relation between CEF levels and magnetic anisotropy, we rewrite the local CEF Hamiltonian as 
\begin{align}
  \mathcal{H}^{\rm loc} = \sum_k^5 B_k O_k,
\end{align}
where
\begin{align}
  O_1 &=\frac{1}{\sqrt{2}}
  \begin{pmatrix}
    1 & 0 & 0 & 0 & 0 \\
    0 & -1 & 0 & 0 & 0 \\
    0 & 0 & 0 & 0 & 0 \\
    0 & 0 & 0 & 0 & 0 \\
    0 & 0 & 0 & 0 & 0 \\
  \end{pmatrix}, 
  O_2 =\frac{1}{\sqrt{2}}
  \begin{pmatrix}
    0 & 1 & 0 & 0 & 0 \\
    1 & 0 & 0 & 0 & 0 \\
    0 & 0 & 0 & 0 & 0 \\
    0 & 0 & 0 & 0 & 0 \\
    0 & 0 & 0 & 0 & 0 \\
  \end{pmatrix},\notag
\end{align}
\begin{align}
  O_3 &=
  \begin{pmatrix}
    0 & 0 & 0 & 0 & 0 \\
    0 & 0 & 0 & 0 & 0 \\
    0 & 0 & 1 & 0 & 0 \\
    0 & 0 & 0 & 0 & 0 \\
    0 & 0 & 0 & 0 & 0 \\
  \end{pmatrix},
  O_4 =
  \begin{pmatrix}
    0 & 0 & 0 & 0 & 0 \\
    0 & 0 & 0 & 0 & 0 \\
    0 & 0 & 0 & 0 & 0 \\
    0 & 0 & 0 & 1 & 0 \\
    0 & 0 & 0 & 0 & 0 \\
  \end{pmatrix},\notag
\end{align}
\begin{align}
  O_5 &=
  \begin{pmatrix}
    0 & 0 & 0 & 0 & 0 \\
    0 & 0 & 0 & 0 & 0 \\
    0 & 0 & 0 & 0 & 0 \\
    0 & 0 & 0 & 0 & 0 \\
    0 & 0 & 0 & 0 & 1 \\
  \end{pmatrix}.
\end{align}
We take the principal axis as the $y$ direction by supposing the cartesian coordinate for the A sublattice in Fig.~\ref{Fig:model}(c).
$B_1$, $B_3$, $B_4$, and $B_5$ represents the parameters for the atomic-energy level, while $B_2$ represents the parameter for the hybridization between the $d_u$ and $d_v$ orbitals.

Figures~\ref{Fig:CEF_dependence}(a), (c), and (e) [(b), (d), and (f)] 
represent the $B_2$ dependences of CEF energy levels ($\Lambda'_{xx}, \Lambda'_{yy}$ and $\Lambda'_{xx} - \Lambda'_{yy}$) by setting $B_1=0.5$, $B_3=-2.0$, and $B_5=0.2$. 
We set $B_4 = -2.5$ for Figs.~\ref{Fig:CEF_dependence}(a) and \ref{Fig:CEF_dependence}(b), $B_4=-1.6$ for Figs.~\ref{Fig:CEF_dependence}(c) and \ref{Fig:CEF_dependence}(d), and $B_4 = -1.3$ for Figs.~\ref{Fig:CEF_dependence}(e) and \ref{Fig:CEF_dependence}(f).

In the case of Figs.~\ref{Fig:CEF_dependence}(a) and \ref{Fig:CEF_dependence}(b),
the ground state is occupied by the $d_{xy}$ orbital, which leads to nonzero matrix elements 
$\bra{e}l_x \ket{g}$ ($\bra{e}l_y \ket{g}$) for $\bra{e} = \bra{d_{zx}} (\bra{d_{yz}})$. 
Thus, $B_2$ does not affect both $\Lambda'_{xx}$ and $\Lambda'_{yy}$.

On the other hand, when the ground state is occupied by the $d_{yz}$ orbital, the anisotropy depends on $B_2$, as shown in Figs.~\ref{Fig:CEF_dependence}(c), \ref{Fig:CEF_dependence}(d), \ref{Fig:CEF_dependence}(e), and \ref{Fig:CEF_dependence}(f), since 
$\bra{e}l_x \ket{g}$ ($\bra{e}l_y \ket{g}$) becomes nonzero for $\bra{e} = \bra{d_{u}}, \bra{d_{v}} (\bra{d_{xy}})$. 
In such a situation, energy levels of $-\alpha d_u + \beta d_v (\beta d_u + \alpha d_v)$ and the ratio of $d_u$ in the lower eigenstate affect $\Lambda'_{xx}$. 
When $B_2$ becomes larger, $|\alpha|$ becomes smaller as shown in the insets of Figs.~\ref{Fig:CEF_dependence}(c) and \ref{Fig:CEF_dependence}(e), thereby $\Lambda'_{xx}$ decreases because of $|\braket{d_u|l_x|d_{yz}}| > |\braket{d_v|l_x|d_{yz}}|$. 
On the other hand, $\Lambda'_{yy}$ is independent of $B_2$ as the energy of $d_{xy}$ is constant against $B_2$. 
Depending on the CEF parameters, the sign change of $\Lambda'_{xx}-\Lambda'_{yy}$ occurs, as shown in Fig.~\ref{Fig:CEF_dependence}(f).

\section{Finite-temperature phase diagram in the presence of the exchange interaction between the electric toroidal dipoles}
\label{app:ETD_scf}

\begin{figure}[th]
  \centering
  \includegraphics[width=\linewidth]{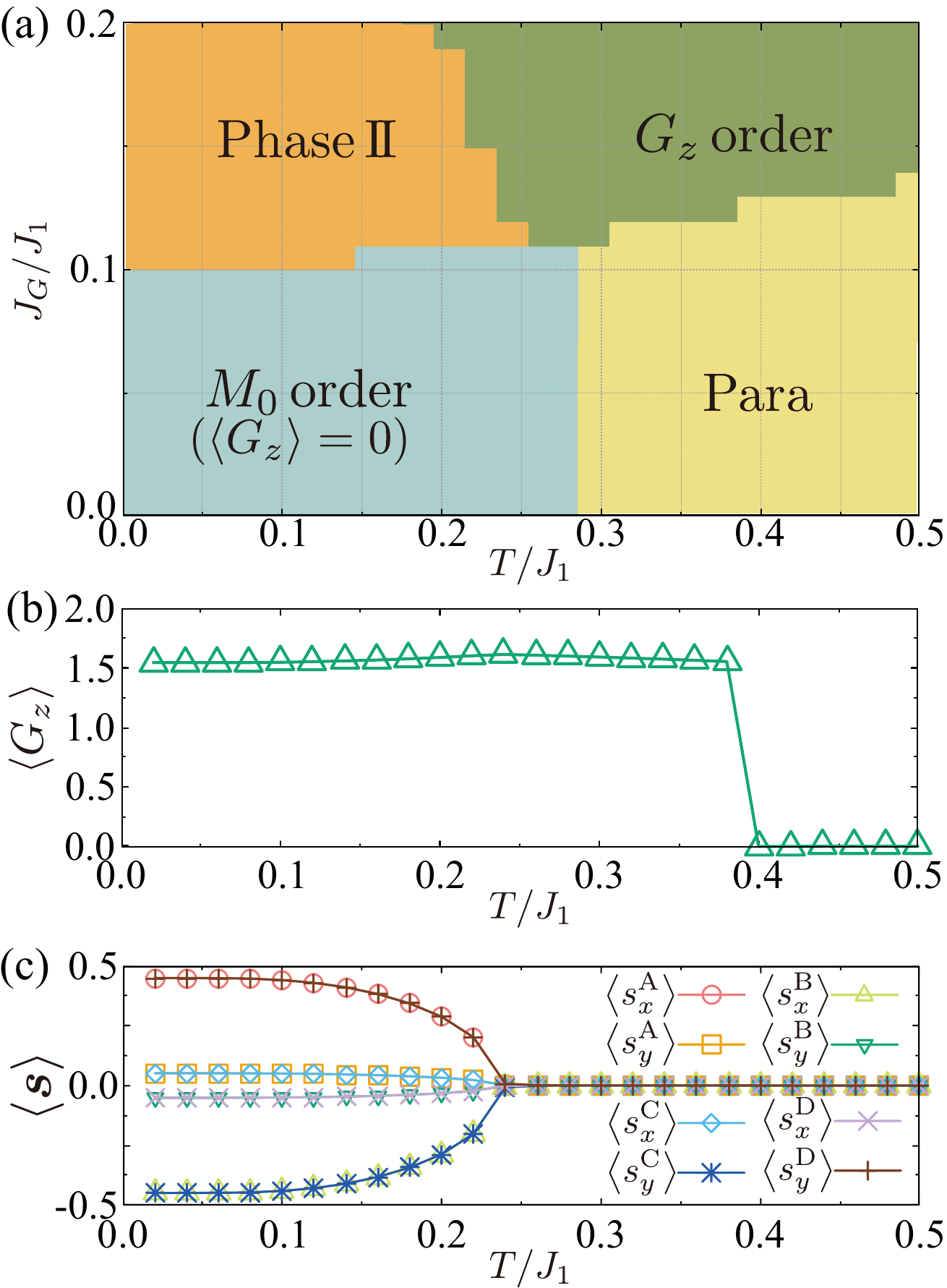} 
  \caption{
    (a) $T$--$J_G$ phase diagram at $\lambda/J_1=0.1$. 
    (b, c) The $T$ dependence of (b) $\braket{G_z}$ and (c) $\braket{s^{\rm A}_x}, \braket{s^{\rm A}_y}, \braket{s^{\rm B}_x}, \braket{s^{\rm B}_y}, \braket{s^{\rm C}_x}, \braket{s^{\rm C}_y}, \braket{s^{\rm D}_x}, \braket{s^{\rm D}_y}$ at $J_G/J_1=0.12$.
    The data is calculated by changing $T$ and $J_G$ with the interval of $\Delta T/J_1=0.02, \Delta J_G/J_1 = 0.01$. 
    The other parameters are the same as those used in Fig.~\ref{Fig:4site_phase}. 
  }
  \label{Fig:phase_ETD_scf}
\end{figure}

In the main text, we deal with the effect of the ETD moment as the one-body molecular-field term $h_G$.
In this Appendix, we introduce the two-body exchange interaction between the ETD moments instead of $h_G$, which is given by 
\begin{eqnarray}
  \label{eq:Gz_scf}
  \mathcal{H}^{\rm G} = -\sum_{\braket{ij}} J^{ij}_G G^i_z G^j_z,  
\end{eqnarray}
where $J^{ij}_G$ is the coupling constant for the nearest-neighbor sites, i.e., $J^{ij}_G=J_G$. 
We apply the mean-field approximation as 
\begin{align}
  \mathcal{H}_{\rm MF}^{\rm G}=
  - J_G \sum_i^{\rm A,B,C,D}\sum_j^{\rm n.n}
  \Braket{G^{i}_z} G^{j}_z + (\rm const.). 
\end{align}
By performing the self-consistent calculations for the four-site cluster model $
\sum_i\mathcal{H}^{\rm loc}_i + \mathcal{H}'^{\rm ex}_{\rm MF}+\mathcal{H}_{\rm MF}^{\rm DM}+ \mathcal{H}_{\rm MF}^{\rm G} $, we obtain the finite-temperature phase diagram against $J_G$ in Fig.~\ref{Fig:phase_ETD_scf}. 
We choose the same model parameters as those in Sec.~\ref{sec:instability} except for $h_G=0$.

Similarly to the results in Sec.~\ref{sec:instability} in the main text, one finds that a sequence of the phase transition occurs for $J_G/J_1 \gtrsim 0.1$; in the case of $J_G/J_1 = 0.12$, the paramagnetic state with $\langle G_z \rangle =0$ turns into the ferroaxial state with $\langle G_z \rangle  \neq 0$ at $T/J_1 \simeq 0.4$, and this state shows a further transition to Phase II at $T/J_1\simeq 0.24$ by decreasing the temperature.
Here, the spin configuration in Phase II is characterized by the linear combination of the magnetic monopole and magnetic toroidal dipole, as discussed in the main text. 
We show the behavior of $\langle G_z \rangle $ and spin moments $\langle s^{i}_{\mu}  \rangle$ for $i=$ A--D and $\mu=x,y$ in Figs.~\ref{Fig:phase_ETD_scf}(b) and \ref{Fig:phase_ETD_scf}(c), respectively. 
Thus, the phase transition from the ferroaxial state to the vortex spin state occurs in a unified way once the ferroaxial moment is induced. 

\section{Result for electric hexadecapole}
\label{app:E16sl}

\begin{figure}[thbp]
  \centering
  \includegraphics[width=\linewidth]{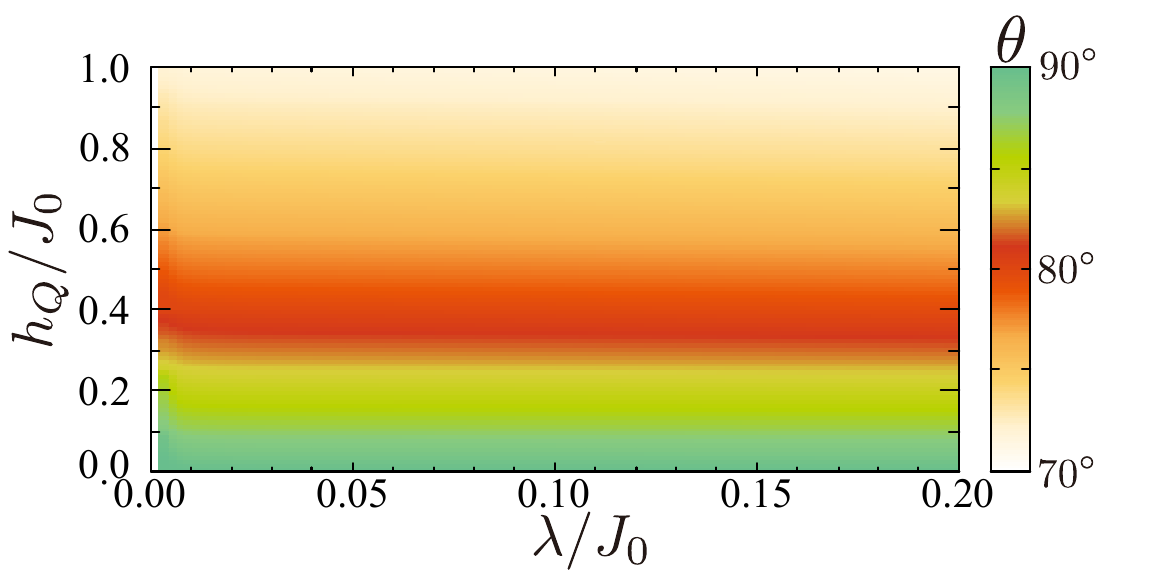} 
  \caption{
  Contour plot of the tilt angle $\theta$ of the spin moments under the $Q^\alpha_{4z}$ molecular field. 
  The other parameters are the same as those used in Fig.~\ref{Fig:1site_theta}.
  It is noted that $\theta$ is indeterminate for $\lambda/J_0=0$ owing to no magnetic anisotropy in the $xy$ plane.
  }
  \label{Fig:E16sl}
\end{figure}

Although we have investigated the ferroaxial ordering under the ETD moment in the main text, other multipoles also lead to the ferroaxial ordering when their irreducible representations are the same as each other.
In the $d$-orbital space, the electric hexadecapole $Q^\alpha_{4z} [\propto xy (x^2 -y^2)]$ is another degree of freedom related to the ferroaxial moments, since $Q^\alpha_{4z}$ belongs to the same irreducible representation as $G_z$ under the $D_{\rm 2h}$ symmetry. 
In this Appendix, we briefly discuss the result for the $Q^\alpha_{4z}$ ordered phase. 
We analyze the single-site model, where we replace the mean-field term $-h_G G_z$ to $-h_{Q} Q^\alpha_{4z}$ in Eq.~(\ref{eq:1site_Hamiltonian}).
The other model parameters are the same as those used in Sec.~\ref{sec:anisotropy}

Figure~\ref{Fig:E16sl} shows the contour plot of the tilt angle $\theta$ by changing $\lambda$ and $h_{Q}$, which is obtained by the self-consistent mean-field calculations.
In contrast to the result in Fig.~\ref{Fig:1site_theta} in Sec.~\ref{sec:anisotropy} in the main text, $\theta$ does not depend on $\lambda$, while $h_Q$ tilts the spin moments from the crystal axis. 
This behavior is attributed to the anisotropic form factor of $\Lambda'_{\mu\nu}$. 
In the presence of $Q^\alpha_{4z}$ without the spin component, the spin Hamiltonian in terms of the $x$ and $y$ spin components is represented as 
\begin{align}
  \label{eq:E16sl_anisotropy}
  \mathcal{H}'_{\rm s}  &= -\lambda^2 \sum_{\mu, \nu = x,y}\Lambda'_{\mu\nu} s_{\mu}s_\nu,
\end{align}
where $\Lambda'_{xx}$, $\Lambda'_{yy}$, and $\Lambda'_{xy}$ can become nonzero in the presence of $h_{Q}$ in contrast to $\Lambda_{xy}$ in Sec.~\ref{sec:anisotropy}.

By diagonalizing $\bm{\Lambda}'$ in Eq.~(\ref{eq:E16sl_anisotropy}), 
one obtains $\theta$ as follows:
\begin{align}
\theta = \arctan \left[
\frac{2 \Lambda'_{xy}}{
\sqrt{(\Lambda'_{xx}-\Lambda'_{yy})^2+4 \Lambda'^2_{xy}}+\Lambda'_{xx}-\Lambda'_{yy}
}
\right].
\end{align}
Thus, one finds that $\theta$ has no $\lambda$
dependence, which is consistent with the numerical results in Fig.~\ref{Fig:E16sl}. 

\bibliographystyle{apsrev}
\bibliography{main.bib}
\end{document}